# Ultra-high Vacuum Deposition of Higher Manganese Silicide Mn$_4$Si$_7$ Thin Films

Rajendra P. Dulal[a], Bishnu R. Dahal, Ian L. Pegg, and John Philip

Department of physics, The Catholic University of America, Washington, DC 20064

The Vitreous State Laboratory, The Catholic University of America, Washington, DC 20064

[a)]Electronic mail: 18dulal@cua.edu

We have successfully grown one of the higher manganese silicides, Mn$_4$Si$_7$ thin films on silicon (100) substrates using an ultra-high vacuum deposition with a base pressure of $1 \times 10^{-9}$ torr. The thickness of the film was varied from 65-100 nm. These films exhibit a tetragonal crystal structure and display paramagnetic behavior as predicted for the stoichiometric Mn$_4$Si$_7$ system. They have a resistivity of $3.321 \times 10^{-5}$ $\Omega$m at room temperature and show a semi-metallic nature.

## I. INTRODUCTION

Higher manganese silicides (HMS) have drawn much attention in recent years due to their potential applications in optoelectronics, thermoelectrics, spintronics, magnetic memories and in magnetic sensors [1-6]. HMS are silicon-rich compounds in the manganese-silicon intermetallic phase diagram. All HMS are referred to as Nowotny Chimney Ladder phases, being derived from the $TiSi_2$ structure [7]. Several HMS phases are reported in the literature, four of them are $Mn_4Si_7$ [8], $Mn_{11}Si_{19}$ [9], $Mn_{15}Si_{26}$ [10], and $Mn_{27}Si_{47}$ [11]. They all display a tetragonal crystal structure with almost equal *a* and long *c* lattice parameters. In the HMS crystal structure, Si atoms form a sublattice by filling intersites in the Mn sublattice, forming a chimney like structure, the Mn sublattice acts as a ladder [12]. Thin films of HMS are generally grown by a solid phase epitaxy or by reactive deposition under high vacuum conditions [13, 14]. But reports show that the secondary phases were present in addition to the HMS phase [15, 16]. It has been reported that the HMS generally exhibit degenerate semiconductor [17-20] behavior. Above 500 K, resistivity starts to decrease exponentially indicating a band gap of about 0.4 eV [17-18, 21]. The Hall effect measurements identify holes to be the dominant charge carriers in HMS systems [17-19, 22]. We have synthesized nanoscale $Mn_4Si_7$ thin films using ultra-high vacuum deposition of Mn and Si, layer by layer on a Si (100) substrate and then annealing at higher temperatures to obtain high quality single phase $Mn_4Si_7$ thin films.

## II. EXPERIMENT

Polycrystalline thin films of $Mn_4Si_7$ were grown using an ultra-high vacuum (UHV) deposition chamber. Cleaned Si (100) substrates were etched with 10% dilute hydrofluoric acid to remove the surface oxide and they were loaded into a UHV chamber with a base pressure of $1 \times 10^{-9}$ torr. Manganese and silicon were deposited layer by layer to a total thickness that was varied from 65 to 100 nm for different batches of nanoscale films. In order to match the stoichiometry of $Mn_4Si_7$, the thickness ratios of manganese to silicon was maintained at 1:2.8. The as-deposited thin films were thermally annealed at 1023 and 1223 K for three hours inside the ultra-high vacuum chamber. The morphology of the thin films was analyzed using a scanning electron microscope (SEM) (JEOL JSM-6300); the crystal structure was investigated by x-ray diffraction (XRD) (Thermo/ARL X'TRA, Cu-$K_\alpha$); and the magnetic measurements were carried-out using a vibrating sample magnetometer (VSM - Quantum Design). Electrical transport measurements were carried out using a Quantum Design physical property measurement System (PPMS) by the four probe resistivity method.

## III. RESULTS AND DISCUSSION

Figure. 1 displays the SEM image of $Mn_4Si_7$ thin films after annealing at 1023 K, which shows an island growth mode. In general, the growth mode of the thin film depends upon the free energy of the substrate surface, interface free energy and surface free energy of the film [23]. The Si and Mn atoms are more strongly bound to each other than with the substrate when the free energy of the substrate surface is low and hence, island structures are formed in the deposited films. Figure. 2 shows the XRD patterns of the thin films annealed at 1023 and 1223 K. For the film annealed at 1023 K, diffraction peaks can be indexed with the tetragonal crystal structure of $Mn_4Si_7$, with lattice parameters, $a$ = 5.525 Å and

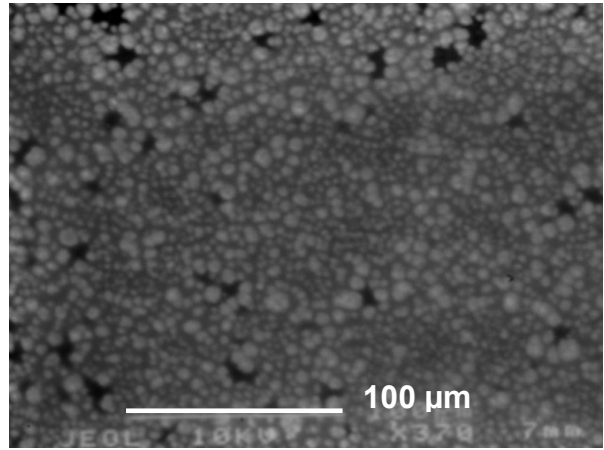

FIG. 1. (Color online) SEM image of 65 nm thin film, after annealing at 1023 K.

$c = 17.463$ Å and space group $P\bar{4}c2$, which is in good agreement with those of bulk samples [8, 24]. The sample annealed at 1223 K showed a mixture of $Mn_4Si_7$ and MnSi phases. Our observation indicates that the single Mn4Si7 phase is formed only when the annealing temperature ranged between 873-1073 K. Secondary phases are observed in the films annealed above 1073 K. It has been reported that the metal silicides may form either by the diffusion of Mn atom in Si layer or diffusion of Si atoms into Mn layer where they react to form silicides [25]. The formation of silicides depends on the difference in electronegativity between Mn and Si atoms. When the temperature is greater than 1073 K, a higher rate of Si-Si bonds breaking occurs apparently. This results in an appreciable Si flow so that the formation of the MnSi phase is possible.

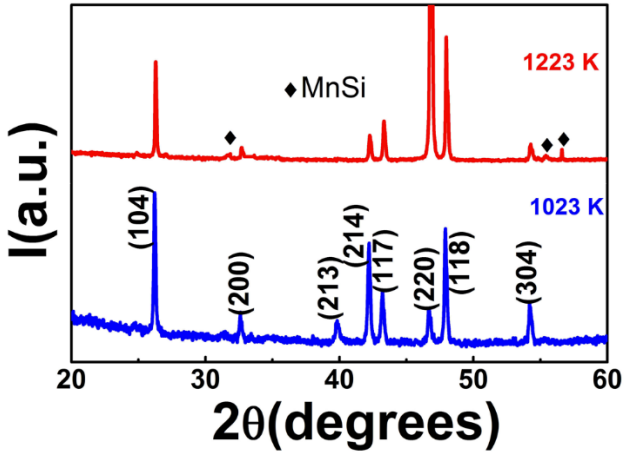

FIG. 2. (Color online) XRD patterns of the thin films annealed at 1023 K and 1223 K.

The magnetic properties of $Mn_4Si_7$ thin film were measured at different temperatures. The magnetic field was applied parallel to the plane of the thin film. Figure. 3 displays the magnetic behavior of the film annealed at 1023 K. From the M-H plot, we observe that the $Mn_4Si_7$ thin films exhibit paramagnetic behavior. Theoretically this system displays zero net magnetic moment and a non-magnetic ground state [20]. The mechanism of magnetism in HMS is similar to that of transition metals where a part of electrons in the localized d-bands transfers to the free electron like s-band, thereby doping holes the in d-band. When the valence electron concentration (VEC) decreases from 14, the density of states at the Fermi level increases rapidly and ferromagnetic ordering should become stable [26]. $Mn_4Si_7$ has VEC exactly equal to 14 so that it does not have any net magnetic moment and no magnetic ordering, which is confirmed by our magnetic measurements.

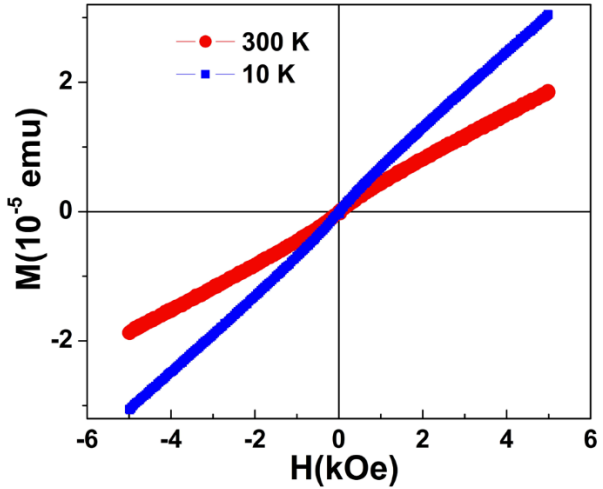

The electrical behavior of $Mn_4Si_7$ thin films was measured from 10 – 300 K. Figure. 4 displays the resistivity as a function of temperature. The $Mn_4Si_7$ thin film exhibits a semi-metallic behavior from 10 K to room temperature. The low temperature resistivity (10-50 K) can be fitted with a $T^3$ dependence as shown in the top inset in Fig. 4. The room temperature resistivity is found to be $3.3 \times 10^{-5}$ Ω m. The resistivity values of $Mn_4Si_7$ films are in the range reported for HMS systems. Reported resistivity values for $MnSi_{2-x}$ have a very large variation of 0.37-100 m Ω cm [27].

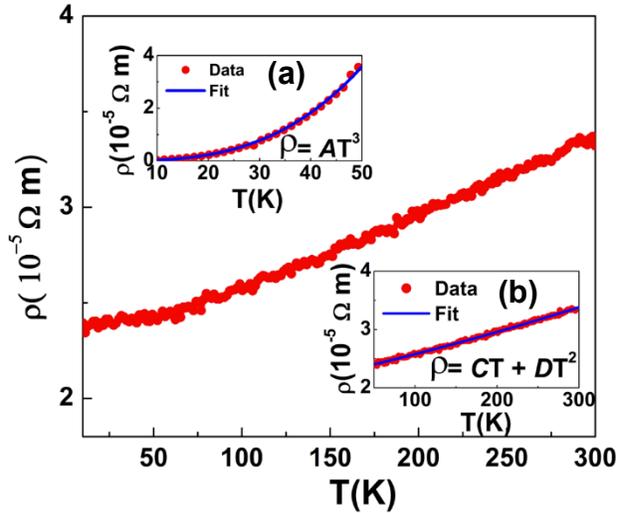

FIG. 4. (Color online) Temperature variation of resistivity for sample annealed at 1023 K. (a) Low temperature resistivity is fitted with a $T^3$ dependence and (b) Resistivity varies as $CT+DT^2$ at high temperature, where $C$ and $D$ are constants.

In semimetals, the overlap of the bottom of the conduction band and the top of the valance band is marginal, therefore, the density of carriers is low and the resistivity is observed to be higher. Resistivity of any material is inversely proportional to the mobility of charge carriers [28]. At low temperature, the mobility of carriers in a $Mn_4Si_7$ thin film is greatly affected by scattering by ionized impurities, defects and grain boundaries [29]. The carrier momentum relaxation is mostly dominated by impurities at low temperature (below 50 K). Since the mean free collision time ($\tau$) is related to temperature, we observed that $\rho = AT^3$ at low temperature (Fig. 4a). However, the effect of impurity scattering decreases with increasing temperature due to the average thermal speed of the carrier being higher. So, acoustic phonon scattering dominates at higher temperature ranges. Hence, the temperature dependence of the resistivity is different from that at low temperature [13]. From our observation, above 50 K, resistivity exhibits a temperature dependence of $CT + DT^2$, where $C$ and $D$ are constants (Fig. 4b).

## IV.  SUMMARY AND CONCLUSIONS

In summary, we have grown higher manganese silicide $Mn_4Si_7$ thin film by ultra-high vacuum deposition. High quality $Mn_4Si_7$ films are obtained only when the films are annealed below 1073 K. Annealing at temperatures greater than 1073 K yields MnSi as a secondary phase together with $Mn_4Si_7$. We have observed that the $Mn_4Si_7$ thin films are paramagnetic and semi-metallic with resistivity, on the order of $10^{-5}$ $\Omega$ m.

## Acknowledgements

This work was supported by the National Science Foundation under ECCS-0845501 and NSF-MRI, DMR-0922997.